\documentstyle[epsfig,preprint,aps]{revtex}
\begin{document} 
\tightenlines
\draft
\preprint{\vbox{\hbox{IFT--P.012/2000}}}
%
%
\title{Can four-fermion contact interactions at one-loop explain the new atomic 
parity violation results?} 
\author{A.\ Gusso}
\address{Instituto de F\'{\i}sica Te\'orica, 
         Universidade  Estadual Paulista, \\  
         Rua Pamplona 145, CEP 01405--900 S\~ao Paulo, Brazil.\\}
\date{\today}
\maketitle

\begin{abstract}
We investigate the possibility that four-fermion contact interactions
give rise to the observed deviation from the Standard Model
prediction for the weak charge of  Cesium, through one-loop
contributions. We show that the presence of loops involving  the
third generation quarks can explain such deviation.
\end{abstract}

\pacs{12.60.-i, 12.60.Rc}

\newpage
\section{Introduction}

For the last fifty years, most of the activity on particle physics
relied on  the use of large particle accelerators. These devices,
allowing the scientists to  break matter down to its most elementary
constituents, have been fundamental in  helping particle physicist
to reveal the secrets of matter. However, besides these
high-energy experiments, low-energy  experiments were also carried
out, giving very important contributions, like the  confirmation of
parity violation in weak interactions. In fact, low-energy 
experiments always played a important role in particle physics. But
now, the  perspectives are that during the first decade of the next
century the importance of  low-energy experiments must increase
significantly. Until LHC collects enough data,  the measurement of
anomalous magnetic moment  of muon \cite{ags} and atomic parity 
violation (APV) in heavy atoms \cite{lan} are going to be a source of
significant new  results \cite{ramsey}.

The measurement of APV in heavy atoms is one of the most important
and ambitious low energy experiments being carried out. The aim is
to  achieve a $0.1 \%$ accuracy in the measurement of the weak charge
of Cesium in the  next few years. Recently a new step was given in
this direction,  the weak charge of Cesium was reported to $0.6 \%$
\cite{experimental},
\begin{equation}
Q_W(^{133}_{55} Cs) = -72.06 \pm (0.28)_{exp} \pm (0.34)_{theor} \; ,
\label{carga}
\end{equation}
We must compare this result with the prediction of the Standard Model
(SM).  Including radiative corrections, it is conveniently expressed
in terms of the oblique  parameters as,  
\begin{equation}
Q_W^{SM}= -72.72 \pm 0.13 - 102 \epsilon^{rad}_3 \; ,
\label{QWSM}
\end{equation}
were the hadronic-loop uncertainty has been included. The value of 
$\epsilon^{rad}_3$ depends on the top quark and Higgs boson mass. For
$m_{top} = 175$  GeV we have \cite{Casalbuoni}
\begin{eqnarray}
\epsilon^{rad}_3 &=& 5.110 \times 10^{-3} \,\, (M_H = 100 {\rm GeV}) \\
\epsilon^{rad}_3 &=& 6.115 \times 10^{-3} \,\, (M_H = 300 {\rm GeV}) \; .
\end{eqnarray}
In the calculations hereafter we assume the $\epsilon^{rad}_3$ given
in Eq. (3). It  is important to stress that our final conclusions are
not going to depend in a  significant way of  $\epsilon^{rad}_3$
dependence on the Higgs mass. Comparing the  theoretical prediction
and the experimental value of $Q_W$ we conclude that
\begin{equation}
Q_W^{exp} - Q_W^{SM} = 1.18 \pm 0.46 \; ,
\label{difference}
\end{equation}
This result implies that the SM prediction and the experimental
result are $2.6 \,  \sigma$ apart. From Eq.\ (\ref{difference}) we
see that the allowed range of  variation for the total new physics
contribution to the weak charge, $\Delta Q_W$, is  
\begin{equation}
0.28 \leq \Delta Q_W \leq 2.08 \; .
\label{interv}
\end{equation}
at $95 \%$ CL. This result is quite interesting. In fact, as noted in
Ref.  \cite{Casalbuoni}, it can be shown that taking seriously the
new result for  $Q_W(^{133}_{55} Cs)$ we can exclude the SM at $99
\%$ CL.

In Ref. \cite{experimental} the  authors see no justification to
believe that such discrepancy originates from  some experimental or
theoretical mistake. They suggest instead that the new  value of
$Q_W$ may have been originated from the presence of some kind of new
physics  beyond the SM. This possibility has already been explored
up  to some extent in Refs. \cite{Casalbuoni,Rosner}, were it is
shown that the observed  deviation in $Q_W$ can be explained by the
presence of a new neutral gauge boson.  Leptoquarks and certain
four-fermion contact interactions can also account for the  present
discrepancy \cite{Casalbuoni}. We point out that all these new 
contributions are at tree-level. No analysis was done considering the
effects of new  physics through one-loop effects. With the intention
of filling partially this gap  we analyze here if four-fermion
contact interactions that do not contribute  at tree-level, can lead
to sizeable contributions to $Q_W$,  through one-loop level 
diagrams.

\section{One-loop effects of four-fermion contact interactions}

Presently, the bounds on new physics are such that the new particles,
if they  exist, must be very heavy. Under these conditions the
effects of these new particles  intermediating  interactions
involving four-fermions can be approximated as contact  interactions.
In the specific case of APV, the contact interactions that can 
contribute at tree-level have the form $g \, (\bar{e}\Gamma  e)(
\bar{q}\Gamma q)$, where $g$ is the coupling constant, $\Gamma$
denotes an  adequate combination of gamma matrices, $e$  is the
spinor for the electron in  the electrosphere, and $q$ corresponds to
the spinor of a quark in the atomic  nucleus. When we want to deal
with one-loop effects we can consider more general  expressions for
the four-fermion interactions. We can consider scalar, vectorial,
and  tensorial interactions involving not only two leptons and two
quarks, as shown above,  but also interactions involving only quarks,
or only leptons. In general, these  interactions can be expressed in
terms of the following Lagrangians  \cite{GonzAndSerg}:
\begin{mathletters}
\label{lag:4}
\begin{equation}
{\cal L}_{\text{scalar}} =  \eta \frac{g^2}{\Lambda^2} 
\left[\bar{\psi}_m \left( V_S^m - i A_S^m \gamma_5 \right) \psi_m
\right] \; 
\left[\bar{\psi}_n \left( V_S^n - i A_S^n \gamma_5 \right) \psi_n \right] \; ,
\label{sca}      
\end{equation}
\begin{equation}
{\cal L}_{\text{vector}} = \eta \frac{g^2}{\Lambda^2} 
\left[\bar{\psi}_m \gamma^\mu \left( V_V^m - A_V^m \gamma_5 \right) \psi_m 
\right] \; 
\left[\bar{\psi}_n \gamma_\mu \left( V_V^n - A_V^n \gamma_5 \right) \psi_n 
\right]\; ,
\label{vec}       
\end{equation}
\begin{equation}
{\cal L}_{\text{tensor}} = \eta \frac{g^2}{\Lambda^2}
\left[\bar{\psi}_m \sigma^{\mu\nu} \left( V_T^m - i A_T^m \gamma_5 \right) 
\psi_m \right]\; 
\left[\bar{\psi}_n \sigma_{\mu\nu} \left( V_T^n - i A_T^n \gamma_5 \right) 
\psi_n \right] \; ,
\label{ten}
\end{equation}
\end{mathletters}
where $\Lambda$ is the energy scale of the effective interaction, 
$V^{m,n}_{S,V,T}$ and $A^{m,n}_{S,V,T}$ are real constants with  $m$
and $n$  being the lepton and quark flavors, and $g$ is the coupling
constant which can  depend on the fermion flavors. The parameter 
$\eta$ can assume the values   $\pm 1$ in order to allow a
constructive or destructive interference  with the  standard
contribution for a given process. Here we have assumed the most
general  four-fermion interactions, in which the new physics present
at high energies must respect only a $U(1)$ symmetry. Such a choice
allow us to  parametrize not only interactions that respect the
$SU(2) \times U(1)$ symmetry  of the SM, but also, and more
accurately, the interesting case of  extensions based on extra $U(1)$
symmetries. 

\begin{center} \begin{figure}
\begin{figure}[ht]
\protect
\epsfxsize=10cm
\begin{center}
\centerline{\epsfig{figure=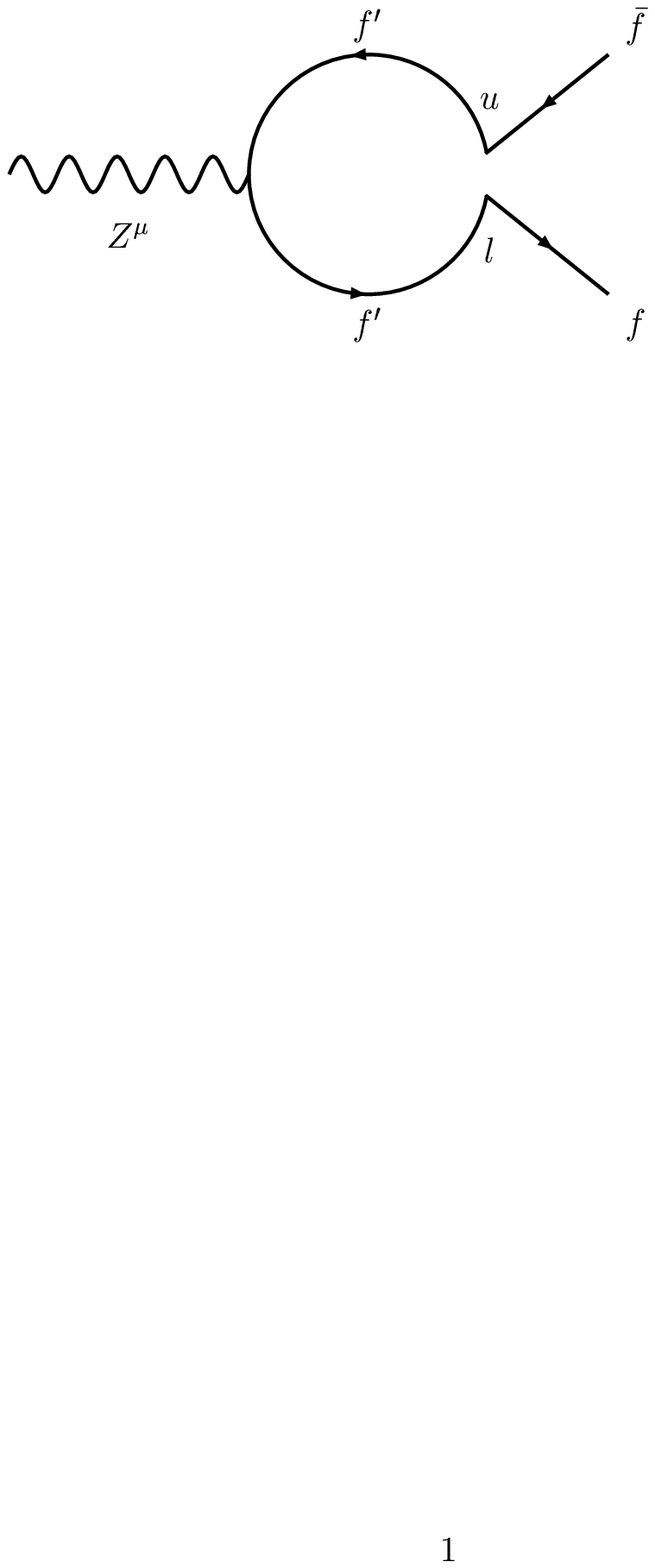,width=0.6\textwidth}}
\end{center}
\end{figure}            
\caption{$s$--channel  diagram.}
\label{s-channel}
\end{figure}
\end{center}

\begin{center} \begin{figure}
\begin{figure}[ht]
\protect
\epsfxsize=10cm
\begin{center}
\centerline{\epsfig{figure=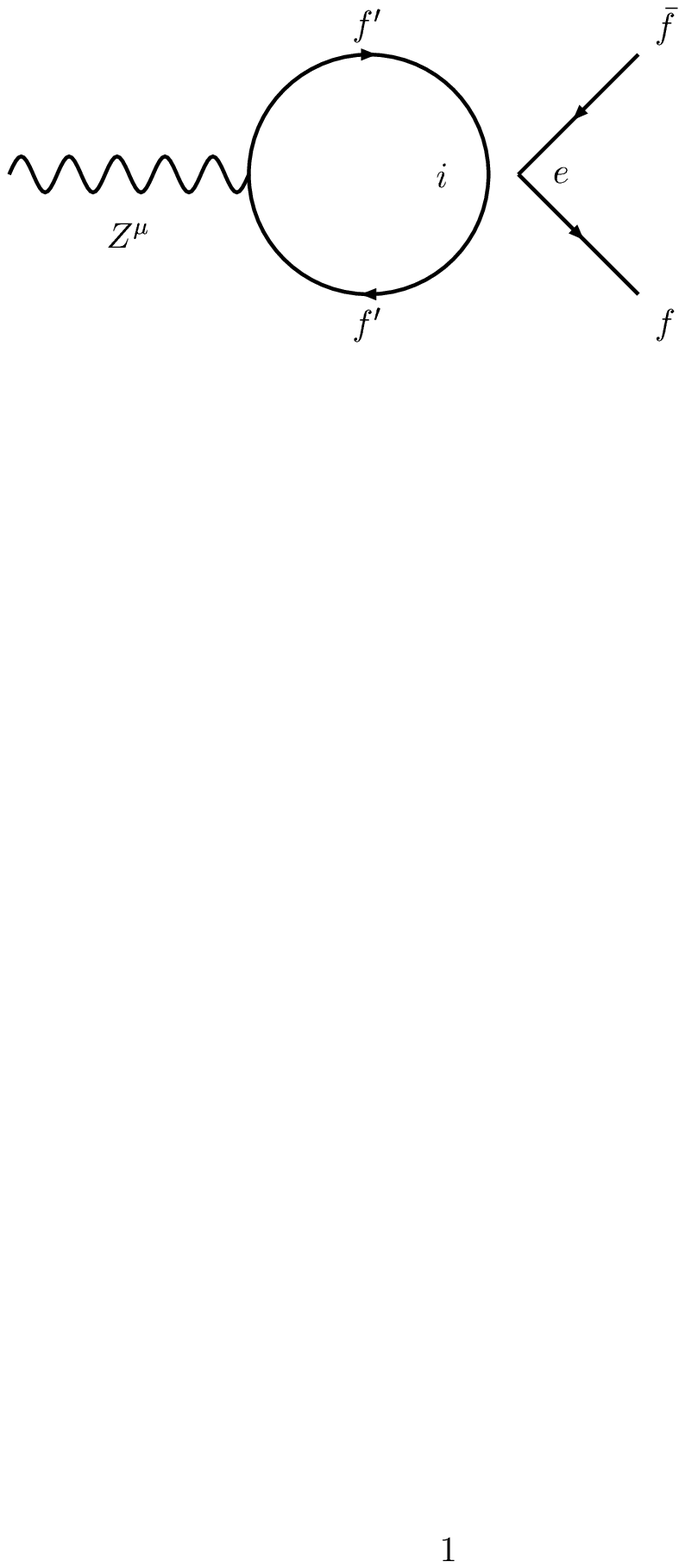,width=0.6\textwidth}}
\end{center}
\end{figure}            
\caption{$t$--channel  diagram.}
\label{t-channel}
\end{figure}
\end{center}

The tensorial and scalar interactions are so severely constrained by
many  experiments \cite{GonzAndSerg,leptoquark} that we will simply
disregard then  hereafter. We consider only the one-loop effects of
the vectorial four-fermion  contact interaction, Eq.\ (\ref{vec}).
The diagrams that contribute to $Q_W$ are  represented in Figs.\
\ref{s-channel} and \ref{t-channel}. In these diagrams the  fermion
$f$ can be either an electron of the electrosphere or a quark of the
nucleus,  and  we allow $f'$  to be any fermion present in the SM.
The only restriction,  obviously, is that the four-fermion
interaction cannot have any significant  contribution at tree-level.
This implies that we do not consider interactions like  $\overline{e}
\gamma e \overline{q} \gamma q$ ($q = u, d$ quarks). The effect of
the  two diagrams is to modify the form factors $F_i$, $i = v, \, a$
in the following $Z$  boson current  
\begin{equation}
J^\mu  = e \; \bar{u}_f (p_1) \;  
\left( F_v \;  \gamma^\mu + 
       F_a \;  \gamma^\mu \gamma_5 \right)
v_f (p_2)  \; . 
\label{current}
\end{equation}
The form factors are functions of $Q^2$, with $Q = p_1 + p_2$.  $F_v$
and $F_a$ are  present at tree--level in the SM
\begin{equation}
F_v^{\text{tree}} \equiv G_V = \frac{1}{2 s_W c_W} 
(T_3^f - 2 \, Q_f \, s_W^2)
\;\; , \;\;\;\; 
F_a^{\text{tree}} \equiv - G_A = - \frac{1}{2 s_W c_W} \,  T_3^f \; ,
\label{g:z}
\end{equation}
where $s_W \, (c_W) = \sin \, (\cos) \theta_W$, $T_3^f$ and is the
third component of  the fermion weak isospin.  The contributions of
the diagrams presented in Figs.\  \ref{s-channel} and \ref{t-channel}
to $F_v$ and $F_a$ have already been evaluated in  Ref.\
\cite{GonzAndSerg}, and are similar to the results of Refs.\ 
\cite{Narison,Mery}.
 
The contribution of the interaction depicted in Eq.\ (\ref{vec}) to
the $s$--channel  is
\begin{eqnarray}
\delta F_v = \eta \frac{g^2}{48 \pi^2 \Lambda^2}  
&&\Biggl\{ \left[6 G_A  M_{f'}^2 - (G_V + G_A) Q^2\right] (
V^l_V + A^l_V)(V^u_V + A^u_V)  
\nonumber \\
&& 
- \left[6 G_A  M_{f'}^2 + (G_V - G_A) Q^2\right] 
(V^l_V - A^l_V)(V^u_V - A^u_V) 
\Biggr\}  
\log\left(\frac{\Lambda^2}{\mu^2}\right) \; ,
\nonumber \\
\delta F_a = - \eta \frac{g^2}{48 \pi^2 \Lambda^2}  
&&\Biggl\{ \left[6 G_A M_{f'}^2 - (G_V + G_A) Q^2 \right] 
(V^l_V + A^l_V)(V^u_V + A^u_V)
\nonumber \\
&&
+ \left[6 G_A M_{f'}^2 + (G_V - G_A) Q^2 \right] 
(V^l_V - A^l_V)(V^u_V - A^u_V)  
\Biggr\} 
\log\left(\frac{\Lambda^2}{\mu^2}\right) \; ,
\label{vec:s}
\end{eqnarray}
and to the $t$--channel,
\begin{eqnarray}
\delta F_v &=& \eta \frac{g^2}{12 \pi^2 \Lambda^2} V_{V}^e
\left[6 G_A^i A_{V}^i M_{f'}^2 - (G_A^i A_{V}^i+ G_V^i V_{V}^i) Q^2 \right]
\log\left(\frac{\Lambda^2}{\mu^2}\right) \; ,
\nonumber \\
\delta F_a &=& -\eta \frac{g^2}{12 \pi^2 \Lambda^2} A_{V}^e 
\left[6 G_A^i A_{V}^i M_{f'}^2 - (G_A^i A_{V}^i + G_V^i V_{V}^i) Q^2 \right]
\log\left(\frac{\Lambda^2}{\mu^2}\right) \; .
\label{vec:t}
\end{eqnarray}
Here, the indexes $u(l)$ denote the coupling constants associated to
the upper  (lower) vertices of Fig.\ \ref{s-channel} and the index
$i$ refer to the coupling  constants of the internal fermion running
in the loop, and $e$ refers to  the external fermion ({\it cf.\ }
Fig.\ \ref{t-channel}). The parameter $\mu$  corresponds to the
characteristic energy scale of the physical process under 
consideration.

\section{Contributions to $Q_W$} 

The one-loop contributions, $\delta F_v$ and $\delta F_a$, are going
to contribute to  the APV in Cesium by modifying the coefficients of
the Lagrangian that conventionally   parametrizes the parity
violating terms in the electron--nucleus interaction 
\cite{MarcSirlinBarger},
\begin{eqnarray}
{\cal L}^{PV} = \frac{G_F}{\sqrt{2}}
\Bigl(&& C_{1u} \; \bar{e} \gamma^{\mu} \gamma^5 e \, \bar{u} \gamma_{\mu} u 
+ C_{2u} \; \bar{e} \gamma^{\mu} e \,\bar{u} 
\gamma_{\mu}\gamma^5 u  
\nonumber \\
&&  + C_{1d}\; 
\bar{e} \gamma^{\mu} \gamma^5 e \,\bar{d} 
\gamma_{\mu} d + C_{2d} \; \bar{e} \gamma^{\mu} e 
\,\bar{d} \gamma_{\mu}\gamma^5 
d + ... \Bigr) \; ,
\label{lagrangian}
\end{eqnarray}
where the ellipsis represent heavy--quark terms $q = s,c,b,t$.  In
heavy atoms, as is  the case of Cesium, coherence effects make the
dominant source of parity violation to  be proportional to the weak
charge given by 
\begin{equation}
Q_W = -2 \left[(2Z + N) C_{1u} + (Z + 2N) C_{1d} \right] \; ,
\end{equation}
where $Z$ and $N$ are the number of protons and neutrons in the 
atomic nucleus,  respectively. So we only need to evaluate the
one-loop effects of four  fermion-contact interactions to the first
and third terms in Eq.\ (\ref{lagrangian}), neglecting all other
contributions. Denoting the new physics contributions to  $C_{1q}$ by
$\delta C_{1q}$, $q = u,d$, we can calculate the effect on 
$Q_W(^{133}_{55} Cs)$, 
\begin{equation}
\Delta Q_W = -376\, \delta C_{1u} - 422\, \delta C_{1d}\; .
\label{deltaQW}
\end{equation}

From the $s$-channel diagram corrections to the $Zee$ vertex of the
electron--nucleus  interaction, it results that
\begin{eqnarray}
\delta C_{1q} =  \eta &&\, N_c \frac{g^2}{4 \pi^2} 
(I_3^q - 2 Q^q s_W^2) I_3^{f'}
\left[ (V_V^l + A_V^l)(V_V^u + A_V^u) 
+ (V_V^l - A_V^l)(V_V^u - A_V^u) \right] 
\nonumber \\ &&\times \left (\frac{M_{f'}}{\Lambda} \right)^2 \log \left 
(\frac{\Lambda}{\mu} \right)^2 \; ,
\label{delszee}
\end{eqnarray}
and from the $t$-channel
\begin{equation}
\delta C_{1q} = \eta \, N_c \frac{g^2}{\pi^2} (I_3^q - 2 Q^q s_W^2 ) 
I_3^{f'} \left ( A_V^e  A_V^{f'}\right ) \left (\frac{M_{f'}}{\Lambda} 
\right)^2 \log \left (\frac{\Lambda}{\mu} \right)^2 \; .
\label{deltzee}
\end{equation}
From the $s$-channel corrections to the $Zqq$ vertex we have
\begin{eqnarray}
\delta C_{1q} = \eta && \, N_c \frac{g^2}{4 \pi^2} I_3^e  I_3^{f'}
\left[ (V_V^l + A_V^l)(V_V^u + A_V^u) - 
(V_V^l - A_V^l)(V_V^u - A_V^u) \right] 
\nonumber \\
&&\times \left (\frac{M_{f'}}{\Lambda} \right)^2 
\log \left (\frac{\Lambda}{\mu} 
\right)^2 \; ,
\label{delszqq} 
\end{eqnarray}
and from the $t$-channel
\begin{equation}
\delta C_{1q} = \eta \, N_c \frac{g^2}{\pi^2} I_3^e I_3^{f'}
\left ( V_V^q  A_V^{f'}\right )  \left (\frac{M_{f'}}{\Lambda} \right)^2 
\log \left (\frac{\Lambda}{\mu} \right)^2 \; .
\label{deltzqq}
\end{equation}
Here $N_c$ denotes the color factor which depends on the number of
quarks present in  each graph. To get Eqs.\
(\ref{delszee})--(\ref{deltzqq}) we have assumed $Q^2 = 0$.  This is
a reasonable assumption because the binding energy of the Cesium
electron   which is considered in the experiments (the outermost one)
is of order of fractions  of an electron-volt.

To proceed with our analysis, the first thing we must do is to choose
the model  or models for the four-fermion interactions. This is done
by choosing the values  of the constants $\eta$, $g$, $V_V$, and
$A_V$ in Eq.\ (\ref{vec}). We are going to  consider that the
four-fermion interactions originate from fermion compositeness. 
Since the exchange of constituents among the fermions takes place in
a strong  interaction regime, we are led  to consider $g^2 = 4 \pi$
(see, e.g. Refs.\  \cite{Narison,Peskin}). In this case, the new
physics scale, $\Lambda$,  corresponds to the compositeness scale.

Initially, we estimate the contributions to $\Delta Q_W$ considering
the present  limits on the new physics scale for contact interactions
involving two  electrons and two other SM fermions 
\cite{GonzAndSerg,PDG,OPAL}. We consider now  only contributions to
the $Zee$ vertex (see Eqs.\ (\ref{delszee}) and  (\ref{deltzee})) and
assume the following choice of parameters,
\begin{eqnarray}
(V_V^l + A_V^l)(V_V^u + A_V^u) &+& (V_V^l - A_V^l)(V_V^u - A_V^u) = 1 \; , 
\nonumber \\
A_V^e  A_V^{f'} &=& \frac14 \; .
\label{model1}
\end{eqnarray}
With this choice the $s$- and $t$-channel contributions are equal. 
We note that  such choice is very reasonable since it is similar to
models like LL, RR, and others  usually considered in the literature
\cite{GonzAndSerg,Peskin,OPAL}. We assume such a  model because what
is really important for our estimates is only the order of  magnitude
of the couplings.  In our calculations we take $\eta$ so that the
final  contribution for $Q_W$ is positive, since negative
contributions are completely  excluded. In Table \ref{table1} we have
the value of $\Delta Q_W$ considering a $b$  quark running in the
loop, calculated separately for each possible quark in the  nucleus
and for the different channels, and  for the sum  of all
contributions.  We  assumed $m_b = 4.5$ GeV, $\Lambda = 3$ TeV, and
$\mu = m_e$, were $m_e$ is the  electron mass. The choice of the
value of $\Lambda$ was based on the results of Refs. 
\cite{GonzAndSerg,OPAL}. We can see that the contributions are quite
small because of  the smallness of the $b$ quark mass. In fact,
because of the dependence on $M_{f'}^2$  in Eqs.\ (\ref{delszee}) and
(\ref{deltzee}) we  obtain even smaller results for lighter fermions
in the loop. The results of the  same calculation considering a $t$
quark in the loop can be found in Table  \ref{table2}. In this case
we used $m_t = 175$ GeV, $\Lambda = 10$ TeV and $\mu =  m_e$. The
choice of the value of $\Lambda$ was based on the results obtained in
Ref.  \cite{GonzAndSerg} which come from the constraints set by the
very precise  measurement of $\Gamma_{\ell \ell}$. In this case, the
results we obtained are  really very interesting. $\Delta Q_W$  is of
the order of magnitude of the expected  correction and even if we
assume that the different contributions in the first two  columns and
rows of Table \ref{table2} interferes destructively instead of 
constructively, we have a result which falls into the interval in Eq.\
(\ref{interv}).

The absence  of good limits on the compositeness scale  of $qqq'q'$
interactions,  involving at least one pair of heavy quarks, does not
allow us to make for the $Zqq$  vertex the same estimates we did for
contributions to $\Delta Q_W$ from  $eeqq$  interactions  present in
$Zee$ vertex.  What we can do is to determine bounds on the  range of
possible values of the compositeness scale compatible with Eq.\ 
(\ref{interv}). We assume that
\begin{equation} 
(V_V^l + A_V^l)(V_V^u + A_V^u) - (V_V^l - A_V^l)(V_V^u - A_V^u) = 1 \;\;\;\; 
\mbox{and} \;\;\;\;
V_V^q  A_V^{f'} = \frac14 \; .
\label{model2}
\end{equation}
in Eqs.\ (\ref{delszqq}) and  (\ref{deltzqq}). This implies that the
$s$- and  $t$-channel contributions are equal. We choose $\eta$ so
that $\delta C_{1u}$ and  $\delta C_{1d}$ are always negative, what
implies  $\delta C_{1u} = \delta  C_{1d}$. Such assumptions allow us
to get the most stringent  bounds on $\Lambda$. In Tables
\ref{table3} and  \ref{table4} we have, respectively, for a bottom
and a top quark in the loop,    the values of $\Lambda$  which give
the  deviations expressed in Eq.\  (\ref{interv}) (we assumed $\mu =
\Lambda_{QCD} \approx 300$ GeV). We evaluated  $\Lambda$ considering 
the contributions  resulting from the $u$ and $d$ quarks present in
the nucleus as we did in  Tables \ref{table1} and \ref{table2}. The
results are shown to one and two channels  contributing. The results
in Table \ref{table3} show us that $Q_W$ is reasonably  sensitive to
the presence of $b$ quark loops. This implies that the presence of 
these loops can possibly explain the observed deviation in $Q_W$. As
expected,  $Q_W$ is very sensitive to the presence of $t$ quark
loops, as can be seen from the results in Table \ref{table4}.

It is worth mentioning that in the previous analysis it is reasonable
to assume  that the new physics scale, $\Lambda$, present in the $s$-
and  $t$-channel diagrams is the same, because the contact
interactions come out of  the exchange of the fermion constituents in
a strong interaction regime. But,  in the case we consider that
massive bosons (e.g. leptoquarks and $Z'$s) are  responsible for the
contact interaction, this generally is not a valid assumption. In 
fact, the $s$-channel diagram can be originated from the exchange of
leptoquarks,  diquarks or dileptons while the $t$-channel diagram
from the exchange of  ordinary massive gauge bosons, like a $Z'$
associated to an extra $U(1)$ gauge  symmetry. We are going now to
consider some implications of the possible  presence of these bosons.

We  note that in the case of the most popular models  for  new
massive vectorial bosons  ($W'$, $Z'$ and leptoquarks) the present
bounds on  their masses always satisfy the condition $M>1$ TeV
\cite{PDG}.  Based on this fact we assume, conservatively, the
existence of four-fermion contact  interactions  with $\Lambda = 1$
TeV, and estimate  the allowed values for the coupling constants.
More exactly, what we do here is  to estimate the allowed values of
$g^2 \left[ (V_V^l + A_V^l)(V_V^u + A_V^u) +  (V_V^l - A_V^l)(V_V^u -
A_V^u) \right]$, $g^2 \left ( A_V^e  A_V^{f'}\right )$,  $g^2 \left[
(V_V^l + A_V^l)(V_V^u + A_V^u) - (V_V^l - A_V^l)(V_V^u - A_V^u) 
\right]$ and $g^2 \left ( V_V^q  A_V^{f'}\right )$ in Eqs.\ 
(\ref{delszee})-(\ref{deltzqq}). We  denote these constants
generically by  $G^2$. Considering that only $\delta C_{1u}$ or
$\delta C_{1d}$ contributes to  $\Delta Q_W$, we obtained the results
shown in Table \ref{table5} for $f'$ being the  top quark. We would
get smaller allowed values in the case the  contributions from the
$s$- and $t$-channel were summed as well as if   $\delta C_{1u}$ and
$\delta C_{1d}$ contributed at the same time.  Notice that  the
numbers in Table \ref{table5} are compatible with the coupling 
constants of the  models in Ref.\ \cite{PDG}. For other lighter
fermions  in the loops, the resulting coupling constants must be
unacceptably large. For  instance, for a $b$ quark it should be of
the order of $4 \pi$, as expected in the  compositeness scenario. 

\section{Final discussion and conclusions}
  
In this article we investigated the one-loop effects arising from 
four-fermion  contact interactions that do not appear in the
Standard  Model. We considered that no new physics contributes at 
the tree-level to the weak charge. This situation arises, for
example, when  the contributions from tree-level
diagrams\footnote{Here we are concerned with  diagrams involving the
electron in the atom electrosphere and the $u$ and $d$ quarks  in the
nucleus. The effects arising from sea quarks are negligible.}
interfere  destructively (see, e.g. \cite{GiustiStrumia}). This allow
us to consider that the  new physics is in a sense universal,
affecting all quarks and leptons and yet not  contributing to $Q_W$
at tree-level. Another possibility is that the new physics  leads to
negligible couplings among light quarks and leptons but sizeable ones
in  interactions involving heavy quarks. 

We estimated the effects of the contact interactions on $Q_W$
analyzing the  contributions to the vectorial and axial form factors.
We  concluded that four-fermion interactions containing the top
quark  can lead to sizeable contributions through $Zee$ and $Zqq$
vertex, when  fermion compositeness is assumed. Four-fermion
interactions that contains the   bottom quark can also lead to
sizeable results through the $Zqq$ vertex if the  compositeness scale
is in the range of few hundred GeV to 1 TeV.

The presence of new massive vectorial bosons, like $Z's$ and
leptoquarks, can also  explain the observed discrepancy in the
measured value of the weak charge of Cesium.  They contribute to
$Q_W$ only at the one--loop level, and can be parametrized by  
four-fermion contact interactions. In this scenario also the top
quark loops are the  responsible for sizeable contributions to $Q_W$.
In fact, it is not surprising that  $Q_W$ is very sensitive to top
quark loops; radiative corrections from the SM   contributes with $1.3
\%$ of the value in Eq.\ (\ref{QWSM}).

We conclude by noting that in spite of the fact that our results are 
only approximate, for the very nature of the calculation of one-loop
diagrams in effective interactions \cite{BurgessLondon}, we expect
that the actual effects  of new physics are not going to be far from
what we have obtained. But we must be  aware that cancellations among
different one-loop diagrams may take place in actual  theories,
leading to non-observable effects. But our results  suggest that
one-loop effects of new physics may contribute significantly to the 
weak charge of Cesium, leading to the observed discrepancy between
SM  prediction and the experimental determination.

\section{Acknowledgment}
The author would like to thank S.\ F.\ Novaes and E.\ M.\ Gregores
for the critical  reading of the manuscript. This work was supported
by Funda\c{c}\~ao de Amparo \`a  Pesquisa do Estado de S\~ao Paulo
(FAPESP).


\begin{table}
\caption{$\Delta Q_W$ for bottom quark in the loop.}
\label{table1}
\begin{tabular}{||c|c|c|c||}
Channel $\backslash$ Quark & $u$ & $d$ & $u + d$ \\
\tableline \tableline
$s$ & 0.003 & 0.002 & 0.005 \\
\tableline
$t$ & 0.003 & 0.002 & 0.005 \\
\tableline
$s +t$ & 0.006 & 0.004 & 0.010 \\
\end{tabular}
\end{table}

\begin{table}
\caption{$\Delta Q_W$ for top quark in the loop.}
\label{table2}
\begin{tabular}{||c|c|c|c||}
Channel $\backslash$ Quark & $u$ & $d$ & $u + d$ \\
\tableline \tableline
$s$ & 0.42 & 0.27 & 0.69 \\
\tableline
$t$ & 0.42 & 0.27 & 0.69 \\
\tableline
$s + t$ & 0.84 & 0.54 & 1.38 \\
\end{tabular}
\end{table}

\begin{table}
\caption{Limits on $\Lambda$, in GeV,  for a bottom quark in the loop.}
\label{table3}
\begin{tabular}{||c|c|c|c|c|c|c||}
$\Delta Q_W$  $\backslash$ Quark & $u$\tablenotemark[1] & $d$\tablenotemark[1] 
& $u + d$\tablenotemark[1] & $u$\tablenotemark[2] & $d$\tablenotemark[2] & $u + 
d$\tablenotemark[2] \\
\tableline \tableline
0.28 & 540 & 580 & 810 & 780 & 830 & 1200 \\
\tableline
2.08 & 180 & 200 & 280 & 270 & 280 & 400 \\
\end{tabular}
\tablenotetext[1]{Only one channel ($s$ or $t$).} \nobreak
\tablenotetext[2]{Both channels.}
\end{table}

\begin{table}
\caption{Limits on $\Lambda$, in TeV,  for a top quark in the loop.}
\label{table4}
\begin{tabular}{||c|c|c|c|c|c|c||}
$\Delta Q_W$  $\backslash$ Quark & $u$\tablenotemark[1] &
$d$\tablenotemark[1]  & $u + d$\tablenotemark[1] &
$u$\tablenotemark[2] & $d$\tablenotemark[2] & $u + 
d$\tablenotemark[2] \\
\tableline \tableline
0.28 & 26 & 28 & 38 & 37 & 40 & 55 \\
\tableline
2.08 & 9.0 & 9.6 & 13 & 13 & 14 & 19 \\
\end{tabular}
\tablenotetext[1]{Only one channel ($s$ or $t$).} \nobreak
\tablenotetext[2]{Both channels.}
\end{table}

\begin{table}
\caption{Limits on $G^2$ for contributions through the $s$-channel.
For the  $t-$channel divide the present values by 4.}
\label{table5}
\begin{tabular}{||c|c|c|c|c||}
$\Delta Q_W$  $\backslash$ Quark & $u$\tablenotemark[1] & $d$\tablenotemark[1] 
& $u$\tablenotemark[2] & $d$\tablenotemark[2] \\
\tableline \tableline
0.28 & 0.039 & 0.059 & 0.026 & 0.026  \\
\tableline
2.08 & 0.29 & 0.44 & 0.19 & 0.19  \\
\end{tabular}
\tablenotetext[1]{From contribution to the $Zee$ vertex.} \nobreak
\tablenotetext[2]{From contribution to the $Zqq$ vertex.}
\end{table}


\end{document}